\documentclass[aps,twocolumn,floatfix,superscriptaddress,longbibliography,nofootinbib]{revtex4-2}
\RequirePackage[sort&compress]{natbib}
\usepackage[english]{babel}
\usepackage{color}
\usepackage{subcaption}
\usepackage{booktabs}
\usepackage{float}

\usepackage{ifthen}
\usepackage{siunitx}
\usepackage{tikz}
\usetikzlibrary{hobby} 
\usetikzlibrary{arrows.meta} 
\tikzset{>={Latex[length=4,width=4]}} 
\usetikzlibrary{calc,intersections,decorations.markings}
\usepackage{siunitx}
\usepackage{xcolor} 

\colorlet{mylightblue}{blue!20}
\colorlet{myblue}{blue!50!black}
\colorlet{mydarkblue}{blue!30!black}
\colorlet{mylightred}{red!10}
\colorlet{myred}{red!50!black}
\colorlet{mydarkred}{red!60!black}
\colorlet{mydarkgreen}{green!30!black}

\tikzset{
  midarr/.style={decoration={markings,mark=at position #1 with {\arrow{stealth}}},postaction={decorate}},
  midarr/.default=0.5
}

\usepackage{amssymb,amsfonts,amsmath}
\usepackage{array}
\usepackage{bm}
\usepackage{xcolor}
\usepackage{comment}

\usepackage[colorlinks = true,
            linkcolor = blue,
            urlcolor  = blue,
            citecolor = blue,
            anchorcolor = blue]{hyperref}

\usepackage{dcolumn}
\usepackage{bm}
\usepackage{float}

\usepackage{xcolor}
\definecolor{RoyalBlue}{HTML}{4169e1}
\definecolor{ForestGreen}{HTML}{228b22}

\usepackage{setspace}
\usepackage{comment}

\usepackage{todonotes}

\newcommand{\ulysse}[1]{{\color{black} #1}}

\begin{document}


\title{Invasion with size-dependent dispersion range}


\author{Ulysse Marquis}
\email{ulysse.marquis@gmail.com}
\affiliation{Fondazione Bruno Kessler, Via Sommarive 18, 38123 Povo (TN), Italy}
\affiliation{Department of Mathematics, University of Trento, Via Sommarive 14, 38123 Povo (TN), Italy}


\date{\today}

\begin{abstract}

    The coalescing colony model provides a minimal framework for biological invasions with long-range dispersion. In its standard formulation, the dispersion range is assumed independent of the size of the invading population. Here, we relax this assumption and consider size-dependent dispersal: a main colony of linear size $r$ emits secondary colonies at distance $r^\mu$, with $0 \leq \mu \leq 1$. We derive the generalized dynamical equations for this extended model and map out the growth phase diagram for the leading order contribution. Depending on $\mu$, the main colony exhibits distinct regimes: linear expansion, power-law growth, exponential regime and finite-time blow-up. We confront these theoretical predictions with a spatially explicit physical model. While the coalescing colony approach correctly captures the scaling of the perimeter, it fails to predict the scaling of the volume. We trace this discrepancy to an effective breakdown of circular symmetry in the morphology of the main colony. Finally, we quantify temporal evolution of the population fraction residing outside of the main colony. The coalescing colony model predicts its decay to~$0$ like a power-law when~$\mu<1$, and a macroscopic amount of the population remains in the secondary colonies at~$\mu=1$. Simulations of the physical model reveal a persistent satellite population not captured by the theory at~$\mu>\mu^*\approx 0.7$. Broadly, our findings highlight how coupling dispersal range to population size fundamentally alters invasion dynamics, with implications for biological invasions, metastatic growth, and urban expansion.
  
\end{abstract}

\maketitle

\section{Introduction}

The spatial growth of populations is often driven by dispersal over distances that exceed the scale of local interactions. Postglacial recolonization~\cite{petit1997}, biological invasions~\cite{clark2001,Shigesada:1997,shigesada1995modeling,Shigesada:2002}, and epidemic spread~\cite{arias2018} all highlight the importance of long-range migration in shaping macroscopic expansion patterns. In plants, seeds can be transported by ants, ballistic ejection, rodents, large vertebrates, wind, or human vehicles, generating a broad range of dispersal distances and mechanisms. Empirical studies have shown that these mechanisms produce highly heterogeneous dispersal kernels, ranging from rapidly decaying, short-ranged distributions to heavy-tailed ones in which rare long jumps play a dominant role \cite{bullock2017}.

The distinction between short- and long-range kernels has major consequences for invasion dynamics. A classical illustration is the so-called Reid’s paradox: the migration rates of trees inferred from paleoecological data following the last glacial maximum are incompatible with purely local dispersal~\cite{Clark1998}. Even extremely rare long-distance events can dominate large-scale expansion, leading to effective invasion speeds far larger than those predicted by diffusion-based models.

Stratified diffusion models, in which local growth is coupled to occasional long-distance colonization events, provide a natural framework to account for such effects \cite{shigesada1995modeling,Shigesada:2002,Shigesada:1997}. Similar nonlocal mechanisms also arise in models of metastatic tumor growth, where secondary tumors are seeded away from the primary mass \cite{Iwata2000,HausteinSchumacher}. Analogous growth processes might arise in the context of urban expansion, as discussed in \cite{marquis2025modelingspatialgrowthcities,marquis2025}.

A simple scaling argument illustrates why heavy-tailed kernels qualitatively modify invasion dynamics. Suppose that the probability that settlers colonize at distance $d$ scales as
\begin{align}
    p(d) \propto \frac{1}{d^\gamma}, \qquad 1 < \gamma < 2,
\end{align}
so that the dispersal kernel is fat-tailed. Consider a large parent colony, and assume that a fraction $f$ of its population participates in colonization. The average dispersion range~$\delta$ over $f$ dispersers is governed by extreme  \ulysse{value statistics and therefore reads}
\begin{align}
    \delta \sim f^{\tfrac{1}{\gamma-1}} .
\end{align}
Importantly, this distance increases with the number of dispersers \ulysse{which might in turn depend on the total population, and therefore on the colony size.} In this regime, invasion is driven by rare, arbitrarily large jumps whose typical length scale grows. Finite-distance or purely diffusive models cannot capture this feedback between population growth and dispersal range.


This observation motivates a generalization of the stratified diffusion models discussed in~\cite{shigesada1995modeling}, and generalized in~\cite{carra2017}. In its original form, the model assumes that {\bf(i)} the emission rate of `propagules' (secondary colonies) scales with the linear size of the main colony, and {\bf(ii)} dispersal occurs at a fixed characteristic distance. While analytically tractable, this framework does not account for scale-dependent dispersal range.

In this paper, \ulysse{we build on Ref.~\cite{carra2017} which relaxed assumption {\bf(i)}, the linear growth of secondary colony rate, and} we introduce a generalization of the Shigesada-Kawasaki (SK) model~\cite{Shigesada:2002, Shigesada:1997} accounting for scale dependent dispersion range. \ulysse{The coalescing colony framework and its mean-field equations are described in section~\ref{sec:model}, along with a computational model.}  In section~\ref{sec:growth}, we map the main growth regimes of the SK model. Sec~\ref{sec:num} presents numerical simulations of the spatial model and compare their behavior with the theoretical predictions made in the previous section.

\section{Invasion models} \label{sec:model}

In this section, we describe firstly an extension of the coalescing colony model, firstly introduced by Shigesada and Kawasaki~\cite{Shigesada:1997}, which can be described analytically. Then, we present a physically grounded model of growth by dispersion, which one can sample using numerical simulations.

\subsection{Coalescing colony model}

\begin{figure}
    \centering
    \includegraphics[width=1\linewidth]{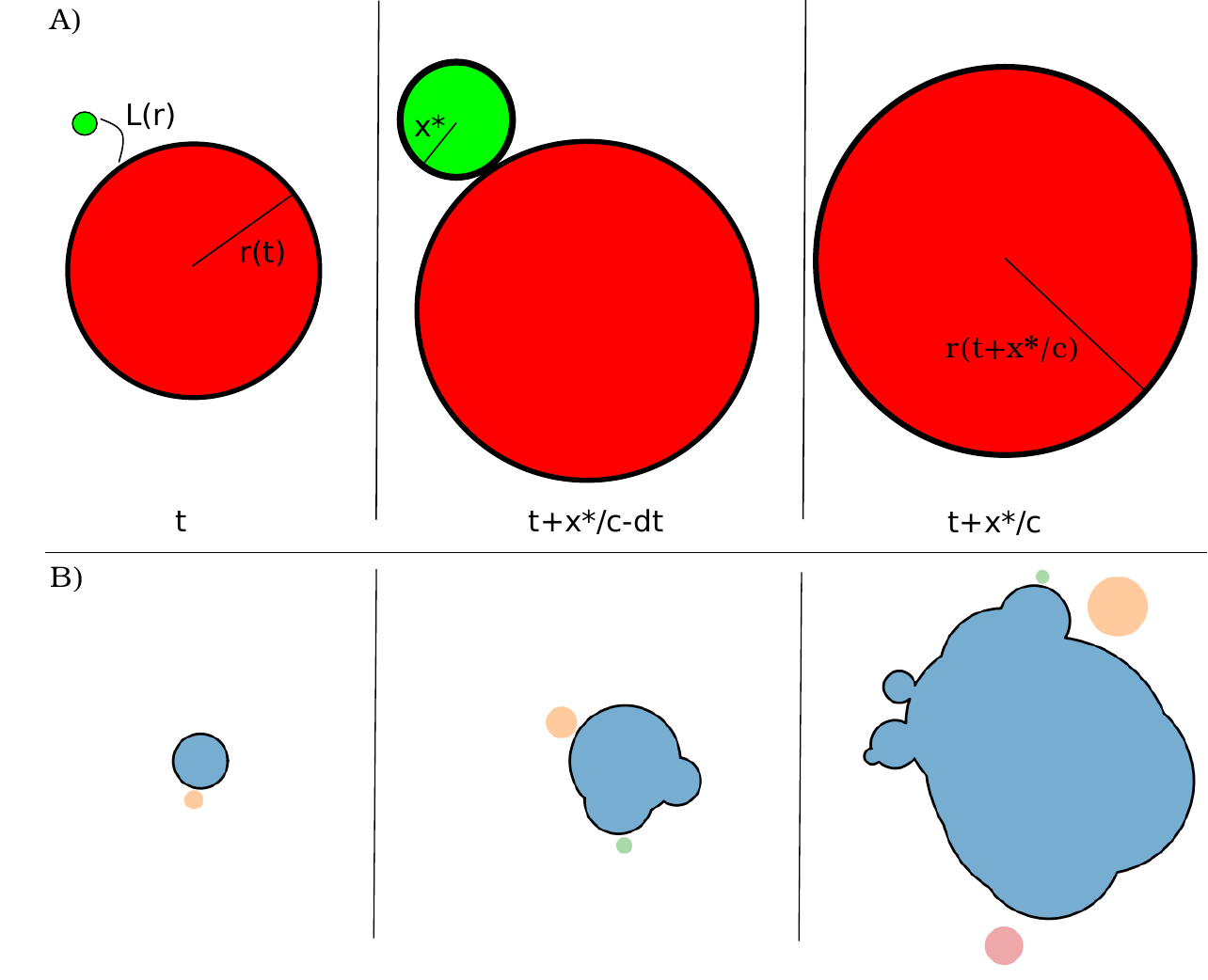}
    \caption{A) Illustration of the coalescing colony model. At rate~$\lambda(r)$, secondary colonies appear at distance~$L(r)$ from the border of the main colony. They grow at radial speed~$c$. When a secondary colony (in green) touches the main colony (in red), it is absorbed instantly and its volume is redistributed isotropically. B) In the physical model instead, colonies still appear at rate~$\lambda(r)$ where~$r$ is an effective radius, but at contact with the main colony, there is simply aggregation. This mechanism can lead to non-trivial morphologies, as shown here. Note that also in both these models, secondary colonies cannot merge between themselves.   }
    \label{fig:models}
\end{figure}

We consider a generalization of the coalescing colony model~\cite{carra2017,Shigesada:1997}, in~$d$ dimensions. A schematic representation is displayed in Fig.~\ref{fig:models}A). A circular primary colony of radius~$r(t)$ grows radially at speed~$c$, and emits secondary colonies at rate~$\lambda(r)$, i.e. a novel colony appears during~$d t$ with probability~$\lambda(r)  \, dt$. Secondary colonies also grow at radial speed~$c$. Such a linear growth approximates (up to a small correction) the Fisher-KPP-like~\cite{Fisher1937, Kolmogorov1937} invading fronts for steep initial conditions, which appear in a wide range of configurations (see~\cite{Marquis_2026} and references therein).

In the original paper~\cite{Shigesada:1997}, three functions were considered for~$\lambda$: 
$$
\begin{aligned}
(1)\quad & \lambda(r) = \text{cst}, \\
(2)\quad & \lambda(r) \sim r, \\
(3)\quad & \lambda(r) \sim r^2.
\end{aligned}
$$
The first case corresponds to population-independent colonization rate, while in the second (resp. third) case, the perimeter (resp. area) of the primary colony controls the dispersion rate. Instead, in this paper, we consider the general case~$\lambda(r) = \lambda_0 r^\theta$ introduced in~\cite{carra2017}. The secondary colonies are initially pointlike and spawn exactly at distance~$L(r) \sim r^\mu$ from the border of the main colony. In previous works,~\cite{shigesada1995modeling, Shigesada:1997, carra2017}, only the case~$L(r) = L_0$ (i.e~$\mu=0$) has been considered. The secondary colonies also grow at speed~$c$, but cannot emit other colonies or merge with an other secondary colony. The only possible merges are with between the main colony and the secondary colonies. This results in an instant absorption of the secondary colony by the main one, with the volume conservation law
\begin{align}
    r^d \leftarrow r^d + \rho^d
\end{align}
where~$\rho$ is the radius of the secondary colony, illustrated in Fig.~\ref{fig:models}A).

Let~$x^*(t)$ be the radius of the colony absorbed at time~$t$ and denote by~$\dot{f} = \frac{df}{dt}$ the temporal derivative of the quantity~$f$. Then, the growth of~$r$ is controlled by the Shigesada-Kawasaki equations
\begin{align} \label{eq:dynr}
    \dot{r} = \begin{cases}
        c & \text{ at } t< t_s \\
        c + \frac{\lambda(r(t-\frac{x^*}{c}))}{c d r^{d-1}} {x^*}^d (c-\dot{x^*}) & \text{at }  t>t_s 
    \end{cases} \,,
\end{align}
where~$t_s$ is the time of the first collision. This equation is derived in~\cite{Shigesada:1997, Shigesada:2002,carra2017} and does not change with the addition of scale-dependent dispersion range. \ulysse{However, the geometric \emph{touching condition} does instead change. In the studied framework, this condition reads:} 
\begin{align} \label{eq:touching}
    L \left( r(t-\frac{x^*}{c}) \right) = r(t) - r \left( t-\frac{x^*}{c} \right) + x^*(t) \,.
\end{align}
\ulysse{This relation can be understood in the following way. Consider a secondary colony of size~$x^*$ at the point of merging with the primary colony. When the former was emitted, at time~$t-\frac{x^*}{c}$, the dispersal range was equal to~$L(r(t-\frac{x^*}{c}))$. This is the left-hand side of~\eqref{eq:touching}, which at the time of contact ($t$) can be equated to the sum of the growth of the size of the main colony since the emission event~$r(t)-r(t-\frac{x^*}{c})$ and of the growth of secondary colony~$x^*(t)$, which is the right-hand side of Eq.\eqref{eq:touching}. Taking the temporal derivative of~\eqref{eq:touching} and collecting the terms proportional to~$\dot{x^*}$ gives}
\begin{align}\label{eq:dynx}
    \dot{x^*} = \frac{\dot{r}'-\dot{r}+\mu r'^{\mu-1} \dot{r}'}{c+\dot{r}'+\mu r'^{\mu-1} \dot{r}'} \, c \,,
\end{align}
where~$r' = r(t-\frac{x^*}{c})$,~$\dot{r'}=\dot{r}(t-\frac{x^*}{c})$.
Equations~\ref{eq:dynr} and~\ref{eq:dynx} constitute the equations describing the average dynamics of the system.

\ulysse{These equations were studied in Ref.~\cite{carra2017} in the case~$L(r)=\text{cst}$ ($\mu=0$). The linear size of the main colony grows linearly for~$\theta\le1$, while above this threshold, a superlinear growth characterized by exponent~$3/(4-\theta)$ is obtained, up to~$\theta<4$.} It is appropriate at this point to discuss a few hypotheses and features of the coalescing colony model and the Shigesada-Kawasaki equation.

\begin{enumerate}
    \item The dispersal range is~$L(r) = L_0 r^\mu$, and for simplicity we set~$L_0=1$. Dispersion exactly at range~$L(r)$ approximates more realistic situations where the dispersion range is stochastic, e.g. there is variance around an average dispersion range..
    \item The instantaneous absorption of colony is unrealistic in most cases but helps for analytical predictions.
    \item The Shigesada-Kawasaki equations describe the growth of an average realisation. These equations are derived using a mean-field assumption: there is a \emph{continous} emission and absorption of colonies, which allows to write a density function for offsprings
    \begin{align}
    \rho(x,t) = \begin{cases}
        0 \quad &\text{when } x > ct \,\\ 
        \frac{1}{c} \lambda(r(t-\frac{x}{c})) \quad &\text{otherwise.}
    \end{cases}
    \end{align}
    The dynamics of~$r$ (Eq.~\ref{eq:dynr}) can be written thanks to this assumption.
    \item We focus our study on the case~$\mu \leq 1$, that is dispersion occurs sublinearly or at most linearly. Beyond that limit, i.e. when the dispersion range scales \emph{super-linearly}, the model loses consistency, as newly formed colonies would eventually disperse over distances larger than the primary one.
\end{enumerate}

\subsection{Physical model}

Secondly, we consider the~\textit{physical model}. The main difference with the coalescing colony model is that absorbed volumed is not redistributed. Hence, it leads to non-circular morphologies, as illustrated in Fig.\ref{fig:models}.  Moreover, the growth rule in direction~$\varphi$ is now
\begin{align}
    r(\varphi,t+dt) = r(\varphi,t) + c \, dt \,.
\end{align}
A similar model has been investigated numerically in~\cite{carra2017}, using the rate exponent~$\theta$ as a control parameter. The authors found that the coalescing colony model does not describe effectively the physics of the physical model when~$\theta \geq 1$, \ulysse{as the effective circularity of the main component is broken, which leads to substantial corrections to the perimeter. In particular, the quantity~$(P/(2\pi r)-1)$, where~$r$ is an effective radius, might be asymptotically non-vanishing.}

\section{Growth regimes for the coalescing colony model} \label{sec:growth}


Now that the coalescing colony model is well defined, we turn to the derivation of the dominant contribution to the growth of~$r$. In~\cite{carra2017}, the authors analyzed in detail the two-dimensional ($d=2$) case and identified distinct scaling regimes depending on the value of~$\theta$. For~$\theta < 1$, the growth is asymptotically linear, up to a logarithmic correction, whereas in the regime~$1 \leq \theta < 4$, the only self-consistent solution is superlinear, with exponent~$\beta = \frac{3}{4 - \theta}$. For~$\theta > 4$, the dynamics becomes even faster, leading to faster than power-law growth, such as exponential behavior or even finite-time blow-up. Importantly, they emphasize two key limitations of the approach: (1) the Shigesada-Kawasaki system is only valid up to a finite time, as it assumes that each absorption event does not immediately trigger another; beyond this regime, numerical results reveal avalanche-like dynamics that are not captured by the equations. (2) Since the early-time behavior is necessarily linear, the system undergoes a crossover toward the asymptotic superlinear regime, reached only after a characteristic minimal time~$t_{\min}$. \\

\begin{figure}
    \centering
    \includegraphics[width=1\linewidth]{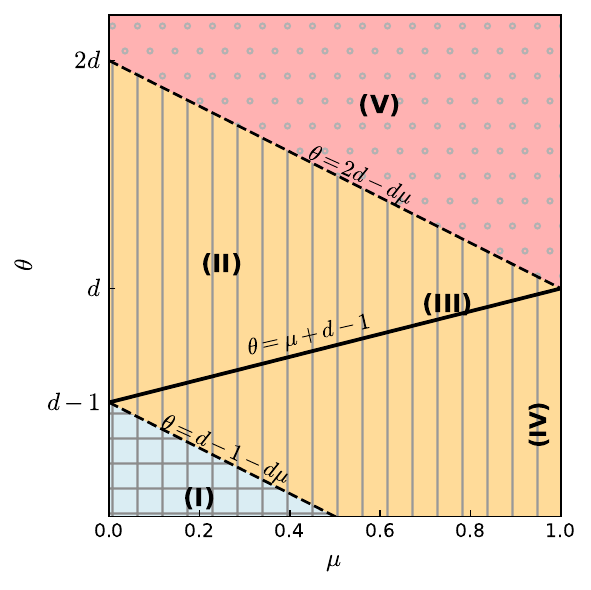}
    \caption{Phase diagram in two dimensions from the system given by equations~\ref{eq:dynr} and~\ref{eq:dynx}. Light blue area ($+$) \textbf{(I)}: linear growth. Orange area ($|$) \textbf{(II)}: superlinear growth with exponent~$\beta = \frac{d+1}{2d - \theta - d\mu}$.  Pink area ($\cdot$) \textbf{(V)}: exponential or finite-time blow-up phase. Brown solid line \textbf{(III)}: critical line, scaling exponent~$\beta = \frac{1}{1-\mu}$. Line~$\mu=1$, $0 \leq \theta < d$ \textbf{(IV)}:~$r \sim t^{\tfrac{d+1}{d- \theta}}$ and~$x^* \sim t$. }
    \label{fig:phasediag}
\end{figure}

The phase diagram mapping the main growth regime is shown in Fig.~\ref{fig:phasediag}. The main regimes are summarized in the following table Tab.~\ref{tab:scaling}.

\begin{table}[h] 
\centering
\begin{tabular}{l|l|l} 
\toprule
Region & $r(t)$ & $x^*(t)$ \\
\midrule
$\theta + d \mu \leq d-1$ 
& $r \sim t$ 
& $x^* \sim t^\mu$ \\

$d-1 < \theta + d\mu < 2d$ 
& $r \sim t^{\beta}$ 
& $x^* \sim t^{\beta(\mu-1)+1}$ \\

$\theta - \mu = d-1$ 
& $r \sim t^{\frac{1}{1-\mu}}$ 
& $x^* \sim \log t$ \\

$\theta + d\mu = 2d$ 
& exponential 
& --- \\

$\theta + d\mu > 2d$ 
& finite-time blow-up 
& --- \\

$\mu = 1,\ \theta < d$ 
& $r \sim t^{\frac{d+1}{d-\theta}}$ 
& $x^* \sim t$ \\
\bottomrule
\end{tabular}
\caption{Summary of growth regimes depending on parameters $\theta$ and $\mu$. 
In the power-law regime, $\beta = \frac{d+1}{2d - \theta - d\mu}$.}
\label{tab:scaling}
\end{table}

For readers interested in the analytical details, the remainder of this section outlines the main steps of the derivation.

\subsection{\underline{Case~$\mu<1$}}
Firstly, we determine whether linear growth can be achieved in the phase diagram, when~$\mu<1$. Assuming~$r \sim v t$, one finds~$\dot{x^*} \sim t^{\mu-1}$ and hence~$x^* \sim t^\mu$. Then, writing
\begin{align}
    r^{\theta-d+1} (t^{\mu})^d \lesssim \text{cst} \,,
\end{align}
we find the condition
\begin{align}
    \theta+ d \mu \leq d-1 
\end{align}
combining dispersion rate and range exponents for growth at linear speed. In particular, when~$\theta+d\mu < d-1$, the asymptotic growth speed is~$c$ while on the line~$\theta+d\mu = d-1$, there is an effective dimension-dependent increase of the speed and the main colony asymptotic speed is~$v>c$.
Secondly, assuming a scaling form~$r \sim t^\beta$ with~$\beta>1$, and plugging it in Eq.~\ref{eq:dynx}, one finds
\begin{align}
    \frac{\dot{x^*}}{c} &= \frac{\overbrace{\dot{r'} - \dot{r}}^{-\frac{x^*}{c} \ddot{r}} + \mu \frac{\dot{r'}}{r'^{(1-\mu)}}}{c + \dot{r'} +\frac{\dot{r'}}{r'^{(1-\mu)} }} \\
    &\sim \frac{x^* t^{\beta-2} + t^{\beta \mu -1}}{t^{\beta-1} + o(t^{\beta-1})}  \label{eq:11} \\
    &\sim \begin{cases}\label{eq:12} 
    t^{\beta(\mu-1)}  & \text{ if }  t^{\beta\mu-1}\gtrsim x^* t^{\beta-2} \, \, \, \, \text{(a)}  \\
     \frac{x^*}{t}  &\text{ if }  x^* t^{\beta-2} \gtrsim t^{\beta\mu-1}   \, \,  \, \,  \text{(b) }
     \end{cases}
\end{align}
Eq.~\ref{eq:11} is obtained via
\begin{itemize}
    \item[$\cdot$] a Taylor expansion on the term~$\dot{r'} -\dot{r}$ which comes from the hypothesis~$t \gg \frac{x^*}{c}$;
    \item[$\cdot$] similarly, ignoring the contribution~$-\frac{x^*}{c}$ and replacing~$r'$ (resp. $\dot{r'}$) by the dominant power-law contribution~$t^\beta$ (resp. $t^{\beta-1}$).
\end{itemize}
Then, two possibilities arise, as shown in  Eq.~\ref{eq:12}. One needs to find which of~$x^* t^{\beta-2}$ or~$t^{\beta\mu-1}$ is dominant. Assume first~$t^{\beta\mu - 1} \gtrsim x^* t^{\beta-2}$. Then~$\dot{x^*} \sim t^{\beta(\mu-1)}$ and~$x^* t^{\beta-2} \sim t^{\beta\mu -1}$ and the assumption is verified. Assume then~$x^* t^{\beta-2} \gtrsim t^{\beta\mu-1}$. Then~$\dot{x^*} \sim  \frac{x^*}{t}$ and then~$x^* \sim t$.~$\frac{\dot{x^*}}{c}-1$ is also of order~$1$ because~$\mu<1$ and hence~$x^* \simeq vt$ with~$v<c$. However, here the touching condition imposes~$\mu=1$ which violates the initial hypothesis. Hence, regime (b) is inexistent in this part of the phase diagram. Note a posteriori that the hypothesis~$\frac{x^*}{c} \ll t$ is consistent:~$x^*$ scales either sublinearly or logarithmically.

One can now derive the values of~$\beta$ from power-counting in Eq.~\ref{eq:dynr}: 
\begin{align}
    \dot{r}-c & \sim r^{\theta-d+1} {x^*}^d \\
    &\sim \begin{cases}
        t^{\beta(\theta-d+1) + d (\beta(\mu-1)+1)}  &\qquad \text{when } \beta \neq \frac{1}{1-\mu}\\
        t^{\beta (\theta-d+1)} (\log t )^d &\qquad \text{else}
    \end{cases} \,.
\end{align}
Ignoring the poly-logarithmic factor, self-consistency gives
\begin{align}
    \beta = \begin{cases}
        \frac{d+1}{2d - \theta-d\mu} & \text{when } \theta-\mu \neq d-1 , \\
        \frac{1}{d-\theta} & \text{when } \theta-\mu = d-1 
    \end{cases} \,.
\end{align}
This regime holds under the condition~$d-1<\theta+d\mu  < 2d$. The lower bound matches the upper bound of the linear regime. Beyond the upper bound~$\theta+d\mu=2d$, we expect exponential and finite-time blow up solutions for the SK equations.

\subsection{\underline{Case~$\mu=1$} } 

At~$\mu=1$, one gets
\begin{align}
    \frac{\dot{x^*}}{c} = \frac{-\dot{r}}{c+2\dot{r'}} \,.
\end{align}
For a power-law growth~$r \sim t^\beta$, this can be re-written
\begin{align}
    \frac{\dot{x^*}}{c} \simeq \left( \frac{1}{\frac{x^*}{ct}-1} \right)^{\beta-1} \,,
\end{align}
which, given that~$\beta>1$, gives~$x^* \simeq c' t $ with~$c' < c$. Then, counting powers in the relation for the dynamics of~$r$, one finds that
\begin{align}
    \beta = \frac{d+1}{d-\theta} \,.
\end{align}
Note the peculiarity of this case: one has~$x^* \sim t$ but~$c-\dot{x^*} \simeq \text{cst}$, which leads to successful power-counting. Because~$t - \frac{x^*}{c} = (1-\frac{c'}{c})t \sim t$ at long times, the factor~$\lambda'$ should have the same scaling exponent as~$\lambda$. A similar analysis at~$\mu>1$ shows that~$\dot{x}^* \to c$ and the aforementioned property is not verified anymore, leading to a degeneracy of the computation.

\subsection{\underline{Validity of the Shigesada-Kawasaki equations}}

As pointed out in~\cite{carra2017}, the Shigesada-Kawasaki equations are valid when the bump provoked by a coalescence event is smaller than the distance between two successive colonies. This condition can be written
\begin{align} \label{eq:validity}
    \left(r^d +{x^*}^d \right)^{1/d} - r < \frac{dc + \frac{d}{dt }r'^\mu }{r'^\theta} \,.
\end{align}
By expanding~$(r^d + {x^*}^d)^{1/d} \simeq r +\frac{1}{d} \frac{{x^*}^d}{r^{d-1}}$, we then have
\begin{align}
    \frac{{x^*}^d}{r^{d-1}} \lesssim \frac{ dc+ \frac{d}{dt} r'^\mu }{r'^\theta} \,.
\end{align}
This formula allows for a rough estimation of the range of validity of the predictions of the coalescing colony model, in particular in the cases where~$x^*$ grows sub-linearly and allows for the approximation~$r' \leftarrow r$. For instance, in the linear phase, one has
\begin{align}
    x^* \sim t^\mu, r\sim vt
\end{align}
and counting the powers of~$t$, one finds that the condition rewrites
\begin{align}
    t^{\overbrace{\theta +\mu d-d+1}^{\leq0} } \leq \text{cst}
\end{align}
which is true asymptotically in the whole linear regime. Secondly, if~$x^* \sim x_0 t^{\beta(\mu-1)+1}$ and~$r \sim r_0 t^\beta$, then
\begin{align}
    t_{\text{SK}} \simeq g ^{1/\nu}
\end{align}
with
\begin{align}
    g = \frac{\mu r_0^{\theta-d-1+\mu}}{x_0^{d[\beta(\mu-1)+1]}} 
     \,,
\end{align}
and~$\nu = \beta(1-\mu)$.
Hence, in this case, the SK equations describe effectively the model up until a certain time, dependent of the parameters of the model. 
Finally, in the regime (III) with~$x^* \sim   \log t$ and~$r \sim t^{\beta}$, the equations are valid up to a time~$t_\text{SK}$ satisfying the self-consistent equation
\begin{align}
    t_\text{SK} \sim  \exp \left[ \mathcal{K} \,  t_{\text{SK}}^{-\tfrac{1}{d}} \right] \,,
\end{align}
where~$\mathcal{K}$ is a constant depending on the parameters of the model.

\section{Numerical investigations} \label{sec:num}

\begin{figure}
    \centering
    \includegraphics[width=1\linewidth]{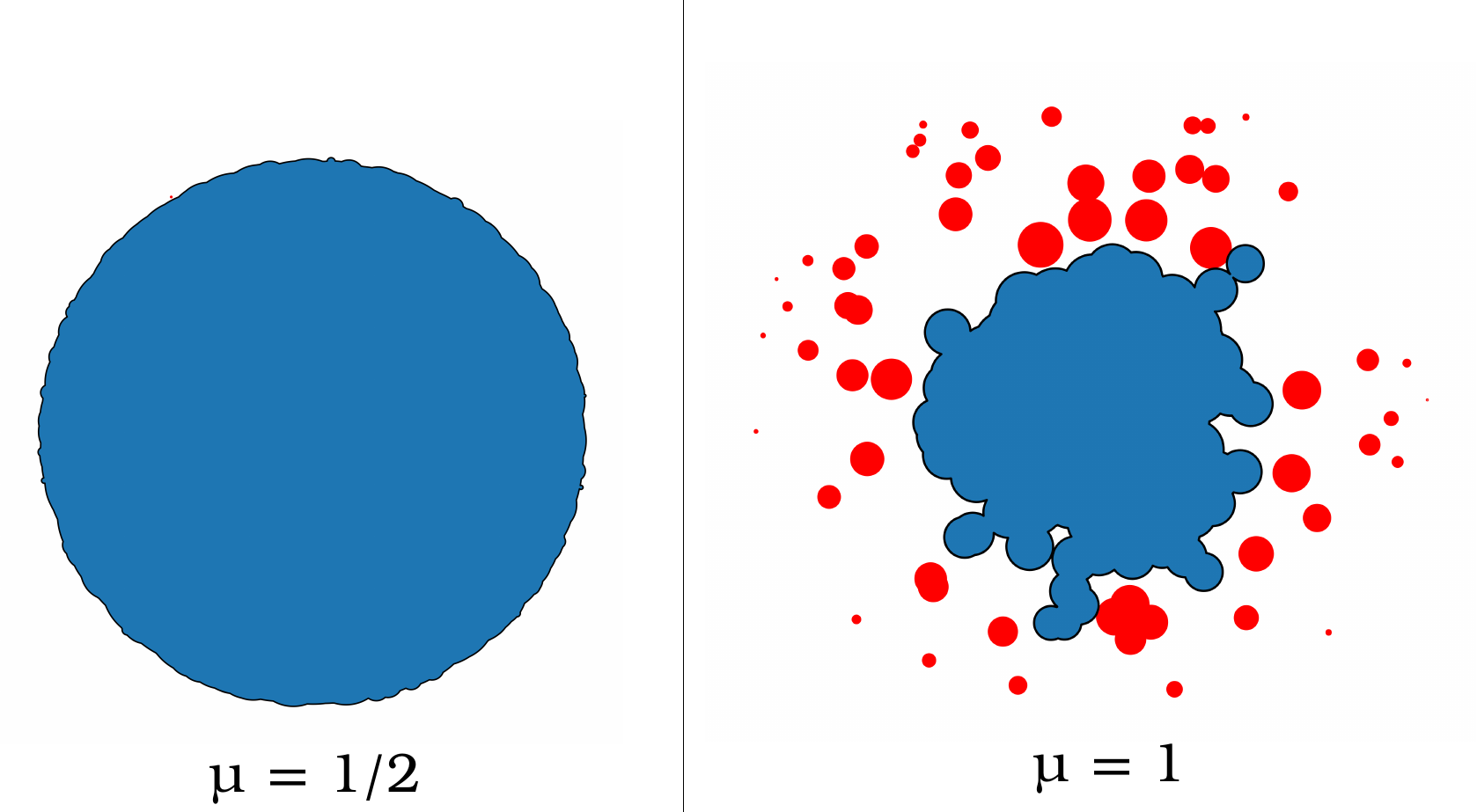}
    \caption{Illustration of the physical model for~$\mu=1/2$ (left) and~$\mu=1$ (right). The main colony is depicted in blue, while the secondary colonies are red. Note the strong non-circularity of the main colony in the right figure.}
    \label{fig:dispersion_illustration}
\end{figure}

In this section, we investigate numerically the physical model described in Section~\ref{sec:model}. Given the difficulty to probe the model at long time when the dispersion rate scales with the size of the main colony (see for instance~\cite{carra2017} which simulates up to times of order~$10^2$), we focus on the case~$\theta=0$ and vary~$\mu$. \ulysse{Verifications of the predictions in two dimensions for~$\theta+d\mu = 1$ (linear case) and superlinear case~$\theta+d\mu=1.5$ are presented in Appendix~\ref{app:a}.}


Fig.~\ref{fig:dispersion_illustration} shows the result of the growth process at large time ($t=10^4$) for~$\lambda(r) = \lambda_0 =  10^{-1}$ inverse unit time. On the left panel, obtained with~$\mu=1/2$, the main colony a) dominates the total area (there is only one secondary colony, of very small size, in the upper left corner) b) circular with small fluctuations of the radius around its average. Instead, for~$\mu=1$ (right panel), a macroscopic fraction of the total invaded area belongs to secondary colonies, and the borders of the main colonies show strong differences with a circle. These observations can be probed quantitatively. Observables are the total volume which is simply the area~$A$ and the perimeter~$P$. In the coalescing colony model, they are both related to the radius whose evolution we track:~$A = \pi r^2$ and~$P = 2 \pi r$.

\begin{figure}[htbp!]
    \centering
    \includegraphics[width=1\linewidth]{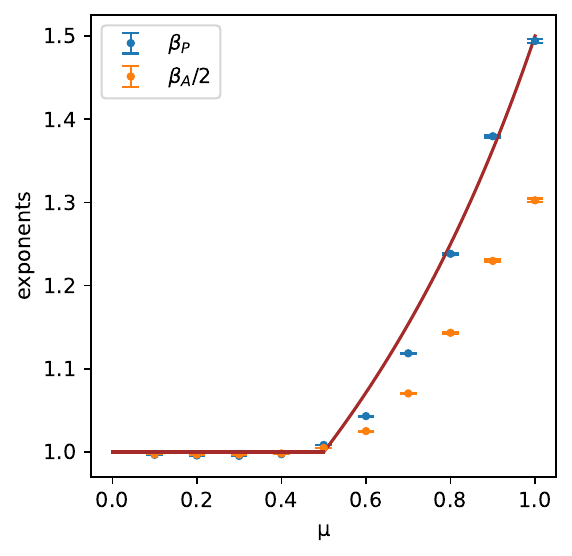}
    \caption{Measured exponents for the perimeter ($\beta_P$) and the area ($\beta_A$). Solid line: predictions of the coalescing colony model. For measuring the scaling exponents, curves were averaged over~$10^3$ realizations. Errorbars :~$95\%$ confidence interval on the exponent.}
    \label{fig:growth}
\end{figure}

Firstly, we measured the growth of the area and the perimeter of the main colony as a function of time, which gives
\begin{align} \label{eq:scaling}
    A \sim t^{\beta_A} ,\qquad P \sim t^{\beta_P} \,.
\end{align}
The measured values of the exponents~$\beta_A/2$ (corresponding to the scaling of an effective radius~$r = \sqrt{A/\pi}$) and~$\beta_P$ are shown in Fig.~\ref{fig:growth}. While the predicted exponents from the coalescing colony model describe well the growth of the perimeter, there are strong corrections to the the scaling of the area, attributable to the non-circularity of the main colony. The breaking of the circularity can be assessed by looking at the breaking of the relation~$P = 2 \pi r$. Introduce the~\textit{circularity}
\begin{align}
    \mathcal{C} = \frac{P}{2 \pi r} - 1 \,,
\end{align}
which is (1) a positive quantity: recall that spheres minimize the surface for a given volume -- the equality case in the isoperimetric inequality -- (2) quantifies the deviations of the perimeter with respect to the circular case. If~$\mathcal{C} \to 0$, the shape is roughly circular. If not, there are strong deviations and if in particular ~$\mathcal{C}$ diverges with time, the growth process shows some fractal characteristics.

\begin{figure}[htpb!]
    \centering
    \includegraphics[width=1\linewidth]{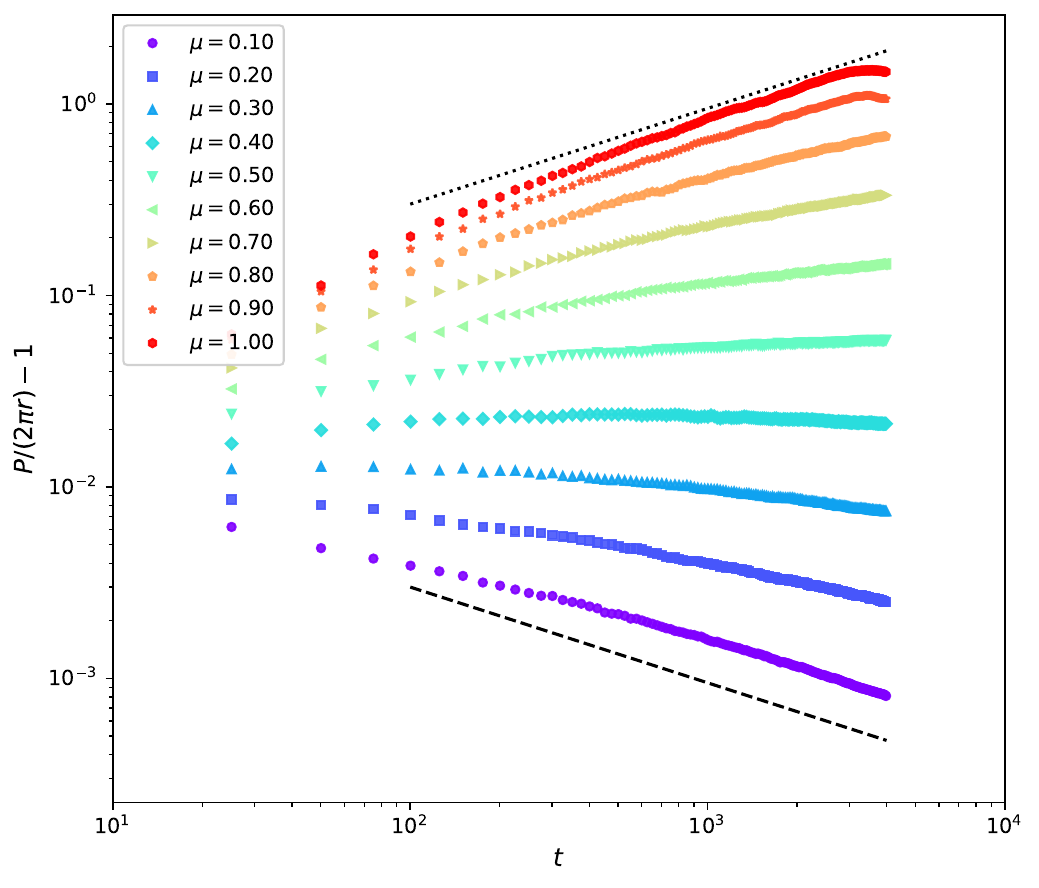}
    \caption{Circularity in function of time. Curves' coloring goes from~$\mu=0.1$ (purple) to~$\mu=1$. For~$\mu<\mu_c \approx 0.4$, the circularity goes to~$0$ like a power-law. Around~$\mu=\mu_c$, the circularity is a constant of time and hence the dependency of the perimeter in the radius is linear but with a correction to the circular factor~$2\pi$. Above~$\mu_c$, the circularity scales with time and there are strong, power law deviations as~$P \sim r^\nu$ (with~$\nu>0$). Curves were averaged over~$10^3$ realizations. Dashed line:~$\sim t^{-0.5}$. Dotted line:~$t^{0.5}$.}
    \label{fig:circularity}
\end{figure}

The evolution of this quantity is displayed in Fig.~\ref{fig:circularity}. Up to~$\mu = \mu_c \approx 0.4 $, the circular approximation holds well.~$\mathcal{C} \sim t^{-\nu}$, with~$\nu\approx1/2$ when~$\mu \to 0$, while at~$\mu \to \mu_c$, the correction seems constant in time, and
\begin{align}
    P \simeq 2\pi (1+\kappa)r \,,
\end{align}
with a linear relation between~$P$ and~$r$. Beyond~$\mu_c$, there are power-law corrections, leading to
\begin{align}
    P \sim r^{1+\nu} \sim A^{\tfrac{1+\nu}{2}} \,,
\end{align}
with~$0 < \nu < 1/2$. The exponent~$\nu$ can be interpreted as a fractal dimension. \ulysse{Our results show that, even at ~$\theta=0$, the phenomena of circularity breaking is observed at large enough values of~$\mu$. Remarkably, the corrections to circularity, quantified by the scaling of~$\mathcal{C}$, even prove to be \emph{superlinear} at~$\mu>\mu_c$ -- this behavior has not been observed when~$\mu=0$ and~$\theta$ is varied~\cite{carra2017}. Indeed, while in the framework of Ref.~\cite{carra2017}, the size of absorbed colonies remain bounded, the growth of the dispersal range induces power-law increase of the size of absorbed colonies (quantified by the scaling of~$x^*$), which amplifies the breaking of circularity. }

\begin{figure}[htpb!]
    \centering
    \includegraphics[width=1\linewidth]{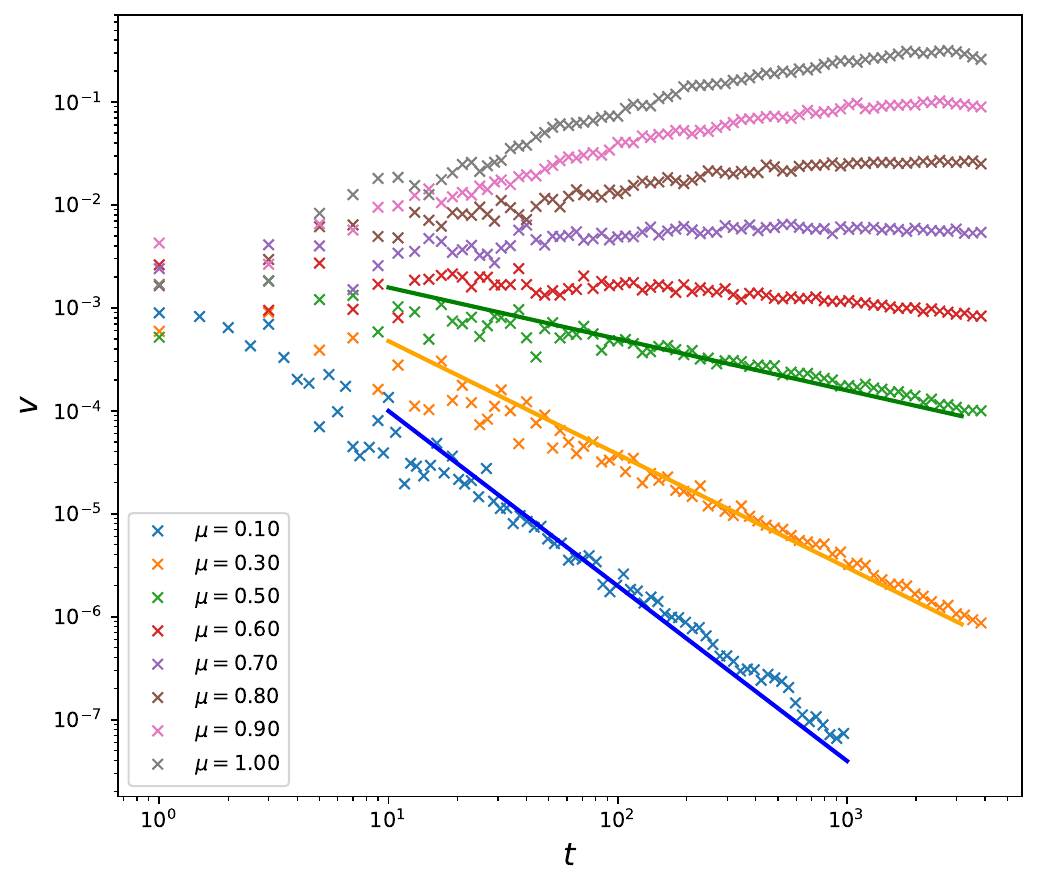}
    \caption{Estimation of the relative secondary colonies volume~$v=\mathcal{V}/V$ (colored crosses). Solid lines: theoretical predictions from Eq.~\ref{eq:volumescaling}. \ulysse{The prefactors yielding the best fits were computed using non-linear regression on the curves at~$t>50$.} Simulations up to time~$t=10^3-10^4$ were run and the shown results are averaged over~$10^2$ realizations.}
    \label{fig:volume}
\end{figure}

Secondly, we assess the relative population in satellite colonies. A simple analysis tells us that in the coalescing colony model, the density of colonies of size~$x$ is~$\rho(x,t) \sim  \lambda(r(t-\frac{x}{c}))$\footnote{In the context of kinetic roughening via random deposition of polydisperse blobs, the size distribution of the blobs (and in particular, its tail) seems to be relevant~\cite{marquis2026dynamicscalingrareevents}.}. Then the volume occupied by the secondary colonies, \ulysse{denoting~$w_d = \pi^{d/2} /\Gamma(\frac{d}{2}+1)$, where~$\Gamma$ is Euler's Gamma function, reads}
\begin{align}
    \mathcal{V} &= \int_{0 \leq x \leq x^*} \rho(x,t) \,w_d \, x^d \, \text{d}x \, \\
    & \sim \begin{cases}
        t^{\mu (d+1) + \theta} \qquad &\text{(I)} \\
        t^{\beta \theta + (d+1)(\beta(\mu-1)+1)} \qquad &\text{(II)} \\
        (\log t )^{d+1} t^{\beta \theta} \qquad & \text{(III)} \\
        t^{d+1  +\beta\theta} & \text{(IV)}
    \end{cases} 
\end{align}
while the volume of the main colony is~$V \sim r^d$. Then, the predictions from the coalescing colony model are
\begin{align} \label{eq:volumescaling}
    v = \frac{\mathcal{V}}{V} \sim \begin{cases}
        t^{\overbrace{\theta+(d+1)\mu-d}^{\leq 0}} & \qquad \text{(I)}\\
        t^{-\beta(1-\mu)} & \qquad \text{(II)} \\
        t^{-\beta(d-\theta) } & \qquad \text{(III)} \\
        \mathcal{O}(1)  & \qquad \text{(IV)}
    \end{cases} \,.
\end{align}
In particular, for the regimes~(I-II-III) a power-law decay is predicted. We can probe numerically the physical model and compare the results. Given the disparity between predicted exponents and actual scaling of the volume at~$\mu\geq0.5$ in the physical model, we expect corrections to the predictions. The question is whether the exponent correction is large enough to change qualitatively the asymptotic behavior, and to obtain persistent satellite population at~$0.5<\mu<1$. In Fig.~\ref{fig:volume}, are shown estimations of the relative volume~$v$, compared to the analytical estimations in the cases~$\mu\leq0.5$. The estimated decay exponent coincides with those found in the simulations, confirming that not only the coalescing colony model is able to predict the scaling of~$r$ correctly but also the one of~$x^*$. At~$1>\mu>0.5$, there are corrections to the prediction. The qualitative change of regime, from power-law decay to constant proportion of the satellite population, occurs at~$\mu^*\approx 0.7$. \ulysse{An interesting trail to explore would be to provide an explicit link between persistent population in secondary colonies and circularity breaking.}


\section{Discussion}

In this work, we have generalized the coalescing colony model of Shigesada and Kawasaki to account for size-dependent dispersion range~$L \sim r^\mu$, for~$\mu<1$. This generalization is motivated by invasion processes with fat-tailed dispersion kernels, for which the dispersion range effectively scales with the population. We uncovered a rich phase diagram characterized by linear, power-law, and explosive growth regimes. 

Firstly, we revealed that there is a subtle interplay between the rate exponent~$\theta$ and the range exponent~$\mu$. An effective exponent~$\theta + d\mu$ controls both the transition from linear growth to power-law growth when~$\theta+d\mu = d-1$ and then from power-law growth to explosive regimes when~$\theta+d\mu=2d$. Moreover, two special regimes appear: when~$\theta-\mu=d-1$, there is power-law growth with exponent~$\frac{1}{1-\mu}=\frac{1}{d-\theta}$, while on the line~$\mu=1$, the exponent is~$\frac{d+1}{d-\theta}$.

Then, with the help of numerical simulations, we analyzed the physical model at~$\theta=0$, which help reveal the strengths and limitations of the Shigesada-Kawasaki approach. On the one hand, the theory correctly predicts the scaling of the perimeter across the explored parameter range. This robustness suggests that the Shigesada–Kawasaki-type approach captures the leading-order kinematics of the advancing front, even when geometric assumptions are relaxed. On the other hand, significant discrepancies arise for the area (volume) scaling in that phase. In particular, mean-field predictions overestimate the scaling exponents. These deviations stem from an effective breaking of circularity: as the dispersal range increases, coalescence events produce strong morphological fluctuations, and the assumption of isotropic volume redistribution becomes invalid. The resulting non-circular shapes induce corrections to the volume scaling that are not captured by the mean-field treatment. This geometric effect has important consequences for the distribution of population between the main colony and its satellites. While the coalescing colony model predicts a power-law decay of the satellite fraction at~$\mu<1$, simulations show instead  a qualitative change at~$\mu^* \approx 0.7$, above which a persistent satellite population remains. The discrepancy can be traced back to the incorrect prediction of the volume scaling in the non-circular regime. In contrast, at~$\mu \leq0.5$ the theory and simulations agree both quantitatively and qualitatively. In summary, introducing a size-dependent dispersal range profoundly enriches the phenomenology of stratified diffusion models. While the coalescing colony framework successfully captures the scaling of front propagation across a broad region of parameter space, its limitations become apparent when geometric effects dominate. This work therefore delineates both the predictive power and the boundaries of validity of mean-field coalescence approaches to invasion dynamics, and open doors to further studies on dispersal models, which relevance ranges from ecological invasions to tumor growth and urban sprawl studies.


\subsection*{Acknowledgements}

UM acknowledges Marc Barthelemy for interesting conversations and insightful comments and Riccardo Gallotti for comments on the manuscript.


\subsection*{Data Availability Statement}

This study does not involve any external datasets. No data were generated or analysed in support of this research.

\subsection*{Funding information}

This research received no external funding.




\markboth{}{}

\bibliography{biblio_ulysse}

\appendix

\renewcommand{\thefigure}{A\arabic{figure}}
\setcounter{figure}{0}

\renewcommand{\thesection}{A\arabic{section}} 
\setcounter{section}{0}   

\clearpage


\appendix

\section{Verification of exponent prediction at~$\theta>0$} \label{app:a}

\begin{figure}[htbp!]
    \centering
    \includegraphics[width=\linewidth]{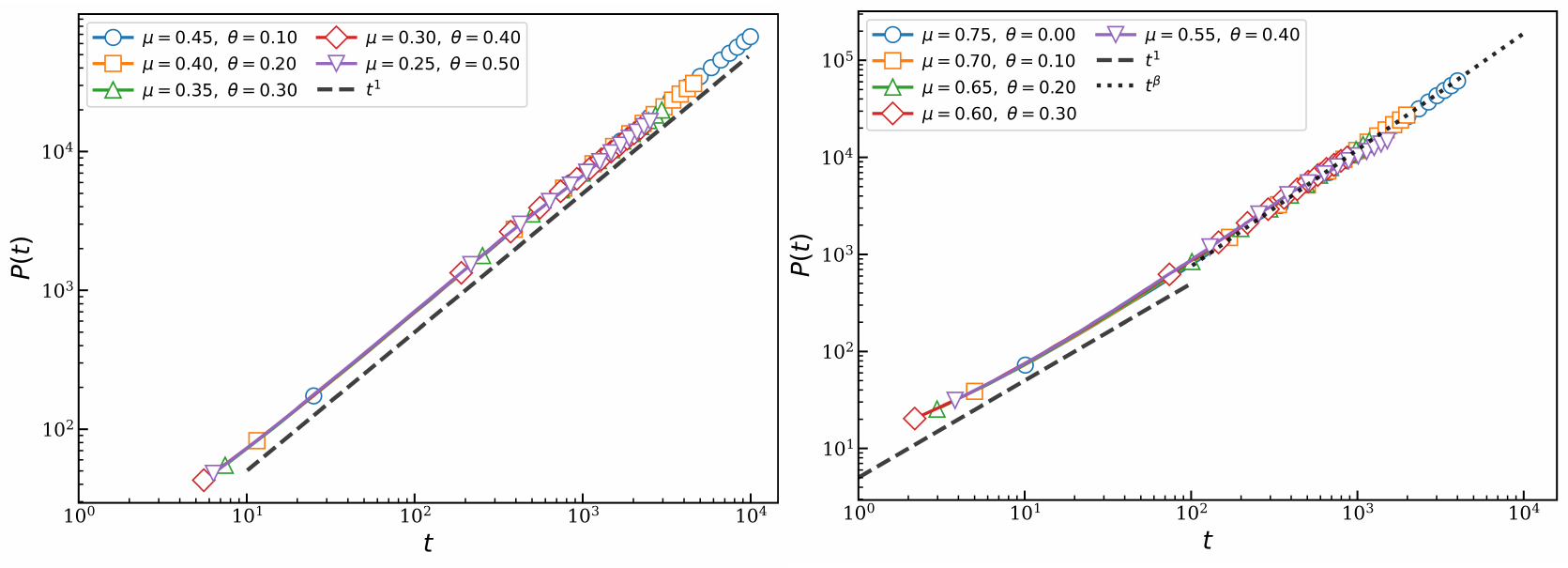}
    \caption{Growth of the perimeter, that is used as a proxy for the radius, as a function of time, for pairs of exponents such that~$\theta+2\mu=1$ (left panel) and~$\theta+2\mu=1.5$ (right panel). The exact values of parameters are shown in the legend. The fit of the intercept of the power-law in the right panel was done using non-linear regression. Each point was computed by averaging over~$10^2$ simulations.  }
    \label{fig:check_exponents}
\end{figure}

\ulysse{In this Appendix, we present numerical verifications of the predictions presented in Section~\ref{sec:model}. We chose pairs of parameters~$(\theta,\mu)$ such that~$\theta+2\mu=C=\text{cst}$, which, following our predictions, should give similar growth behavior. Constants~$C=1$, corresponding the boundary between phase (I) and (II) and predicting a linear behavior, and~$C=1.5$, in the bulk superlinear phase with exponent~$\beta=6/5=1.2$. The numerical results are displayed in Fig.~\ref{fig:check_exponents}. Asymptotically, the data agree remarkably with the predictions. }

%

\end{document}